# Synchronization of a class of master-slave non-autonomous chaotic systems with parameter mismatch via sinusoidal feedback control

Jianping Cai[1,*], Mihua Ma[1], Xiaofeng Wu[1,2]

[1]*Department of Mathematics, Zhangzhou Normal University, Zhangzhou 363000, China*

[2]*Guangzhou naval Academy, Guangzhou 510430, China*

[*]Corresponding author. Email address: mathcai@hotmail.com

Tel: 86-596-2527285, Fax: 86-596-2527012

**Abstract** In this paper we investigate a master-slave synchronization scheme of two n-dimensional non-autonomous chaotic systems coupled by sinusoidal state error feedback control, where parameter mismatch exists between the external harmonic excitation of master system and that of slave one. A concept of synchronization with error bound is introduced due to parameter mismatch, and then the bounds of synchronization error are estimated analytically. Some synchronization criteria are firstly obtained in the form of matrix inequalities by the Lyapunov direct method, and then simplified into some algebraic inequalities by the Gerschgorin disc theorem. The relationship between the estimated synchronization error bound and system parameters reveals that the synchronization error can be controlled as small as possible by increasing the coupling strength or decreasing the magnitude of mismatch. A three-dimensional gyrostat system is chosen as an example to verify the effectiveness of these criteria, and the estimated synchronization error bounds are compared with the numerical error bounds. Both the theoretical and numerical results show that the present sinusoidal state error feedback control is effective for the synchronization. Numerical examples verify that the present control is robust against amplitude or phase mismatch.

**Keywords:** chaos synchronization; robustness; parameter mismatch; sinusoidal feedback control

## 1. Introduction

A lot of efforts have been devoted to the study of synchronization due to its wide-scope potential applications in secure communication,[1-4] chemical and biomedical sciences,[5-6] life science,[7] electromechanical engineering[8-9] and other subjects. Since Pecora and Carroll applied replacing variable control to synchronize two chaotic systems,[10] a wide variety of control techniques have been proposed for the synchronization of chaotic systems, such as linear feedback control,[11-13] adaptive control,[14-16] active control,[17-19] sinusoidal feedback control,[20-22] and others.[23-25] It is well known that even though the identical master and slave systems are placed in a synchronization scheme, parameter mismatch between two systems often occurs because of inevitable perturbation in operations, which can destroy the synchronization.[26-28] Thus, it is particularly important to investigate robust synchronization in the presence of parameter mismatch.[29-44] Recently, synchronization of a coupled chaotic semiconductor lasers in the presence of parameter mismatch have been studied.[29-31] Suykens *et al* applied standard plant form of synchronization scheme to study robust synchronization of master and slave Lur'e systems with parameter mismatch via linear state error feedback control, where the master system is subjected to an external input.[32] Synchronization of Lur'e systems with parameter mismatch was also discussed.[33-34] Two-dimensional lattices of diffusively coupled chaotic oscillators with parameter mismatch were studied by using the scalar diffusive coupling and it was shown that in lattices of Lorenz and Rössler systems the cluster synchronization regimes are stable and robust against up to 10%-15% parameter mismatch.[35] It was found in Ref.36 that only negative conditional or transverse Lyapunov exponents can not guarantee synchronization of two chaotic systems with parameter mismatch. Refs.37-39 investigated synchronization of chaotic Chua's systems in the presence of parameter mismatch. Clearly the contributions mentioned above are significant theoretically and practically. However, most of these papers concerned synchronization of autonomous systems[29-43] and the obtained synchronization criteria were in the form of linear matrix inequalities (LMI),[32-33, 37-38] most of which must be solved by means of computer algebraic systems, such as *Matlab*. On the contrary, synchronization criteria in the form of algebraic inequalities can reveal the relationship between the criteria and system parameters, so it is more convenient to design and analyse the synchronization controller. Synchronization criteria for two-dimensional horizontal platform systems with phase mismatch via linear state error feedback control was derived in the form of algebraic inequalities.[44] The authors investigated synchronization criteria of two second-order non-autonomous chaotic oscillators coupled by a linear state error feedback control, where parameter mismatches may occur in the external excitations or parametric excitations.[45] To the best of our knowledge, research on chaos synchronization of non-autonomous chaotic systems of higher dimension with parameter mismatch has been received little attention up to now. Nevertheless, more and more non-autonomous chaotic systems have been found in engineering and physics.[46-48] Especially, high-dimensional non-autonomous chaotic systems are often met with in practice, for examples, a three-dimensional gyrostat system,[22] and a four-dimensional loudspeaker system.[46]

In this paper, we investigate a master-slave synchronization scheme of two n-dimensional non-autonomous chaotic systems coupled by a sinusoidal state error feedback controller, where parameter mismatch exists between the external harmonic excitation of master system and that of slave one. One of the advantages of sinusoidal controller is that it always produces smooth and bounded output regardless of the magnitude of state error variables.[20] Parameter mismatch in this paper implies that amplitude, frequency and phase can be all or partly different between the external harmonic excitation of master system and that of slave one.

In the following, we firstly present a master-slave synchronization scheme and estimate the synchronization error bound by the Lyapunov direct method and an inequality technique. Secondly, the Gerschgorin disc theorem is used to derive some synchronization criteria of algebraic inequalities for the scheme. We also establish the relationships among the estimated synchronization error bound, parameter mismatch and coupling coefficients. The relationships reveal that the synchronization error can be controlled as small as possible by increasing the coupling strength or decreasing the magnitude of mismatch. Finally, simulations on a three-dimensional gyrostat system verify the effectiveness of the obtained synchronization criteria. The difference between the estimated synchronization error bounds and the numerical error bounds is also illustrated. Both theoretical and numerical results show that the present sinusoidal state error feedback control is effective for the synchronization. Numerical examples verify that the present control is robust against amplitude or phase mismatch.

## 2. Synchronization scheme

Consider a master-slave synchronization scheme for two non-autonomous chaotic systems with different external harmonic excitations coupled by a sinusoidal state error feedback controller as follows:

master system: $\dot{x} = A(t)x + F(x) + m(t)$, (1)

slave system: $\dot{y} = A(t)y + F(y) + \tilde{m}(t) + u(t)$, (2)

controller: $u(t) = (k_1 \sin(x_1 - y_1), k_2 \sin(x_2 - y_2), \cdots, k_n \sin(x_n - y_n))^T$, (3)

where the dot represents the derivative with respect to time $t$, $x = (x_1, x_2, \cdots, x_n)^T \in R^n$, $y = (y_1, y_2, \cdots, y_n)^T \in R^n$, $A(t) \in R^{n\times n}$ is a bounded matrix, $F$ is a continuous nonlinear function, $m(t) = (m_1(t), m_2(t), \cdots, m_n(t))^T \in R^n$ is external harmonic excitation, $m_l(t) = f_l \sin(\omega_l t + \varphi_l)$ (or $m_l(t) = f_l \cos(\omega_l t + \varphi_l)$ ), $l = 1, 2, \cdots, n$ . $\tilde{m}(t) = (\tilde{m}_1(t), \tilde{m}_2(t), \cdots, \tilde{m}_n(t))^T \in R^n$ , $\tilde{m}_l(t) = \tilde{f}_l \sin(\tilde{\omega}_l t + \tilde{\varphi}_l)$ (or $\tilde{m}_l(t) = \tilde{f}_l \cos(\tilde{\omega}_l t + \tilde{\varphi}_l)$), $\tilde{f}_l, \tilde{\omega}_l, \tilde{\varphi}_l$ can be different from $f_l, \omega_l, \varphi_l$, respectively, and $k_1, k_2, \ldots, k_n$ are coupling coefficients. In this paper, we take $m_l(t) = f_l \sin(\omega_l t + \varphi_l)$ and $\tilde{m}_l(t) = \tilde{f}_l \sin(\tilde{\omega}_l t + \tilde{\varphi}_l)$ in the following discussion. The present method can also be applied to study the other situations.

Usually, the synchronization scheme (1)-(3) is difficult to achieve complete synchronization due to parameter mismatch. Therefore a concept of synchronization with error bound is introduced as follows.

**Definition 1** The synchronization scheme (1)-(3) achieves synchronization with error bound $h$ in a region $D$ if for any bounded initial states of systems (1) and (2) in the region $D$, there exist a constant $h>0$ and a $T_0>0$ such that $\|x(t)-y(t)\|<h$ for all $t>T_0$, where $\|\cdot\|$ denote the Euclidean norm.

According to this definition, our control objective is to select suitable coupling coefficients $k_1,k_2,\cdots,k_n$ such that the distance between trajectories $x(t)$ and $y(t)$ of master and slave systems will eventually become less than an error bound $h$ if the initial states $x(0)$ and $y(0)$ lie within a certain region $D$. To this end we need some hypotheses.

**Hypothesis 1** There exists a bounded matrix $M(x,y)$ such that

$$F(x)-F(y)=M(x,y)(x-y), \tag{4}$$

where the elements of $M(x,y)$ are dependent on $x$ and $y$.

Fortunately, such a matrix $M(x,y)$ exists in many practical chaotic systems, such as Lorenz system, Duffing equation as well as the example in Section 6.

## 3 Estimation of synchronization error bound

Defining an error variable $e=x-y$, one can obtain the error dynamical system:

$$\dot{e}=A(t)(x-y)+F(x)-F(y)-u(t)+\Delta m=(A(t)-K(t)+M(x,y))e+\Delta m, \tag{5}$$

where

$$K(t)=diag\{k_1u_1(t),\,k_2u_2(t),\cdots,k_nu_n(t)\},\ \ u_i(t)=\frac{\sin(x_i-y_i)}{x_i-y_i},\ \ i=1,2,\ldots,n,$$

$$\Delta m=m(t)-\tilde{m}(t)=(m_1(t)-\tilde{m}_1(t),m_2(t)-\tilde{m}_2(t),\cdots,m_n(t)-\tilde{m}_n(t))^T\in R^n,$$

$$m_l(t)-\tilde{m}_l(t)=f_l\sin(\omega_l t+\varphi_l)-\tilde{f}_l\sin(\tilde{\omega}_l t+\tilde{\varphi}_l),\ \ l=1,2,\cdots,n.$$

In what follows Lyapunov direct method is used to analysis the stability of the error dynamical system (5) and derive the estimated synchronization error bound. Choose a quadratic Lyapunov function $V(e)=e^T Se$ with the positive definite matrix $S=diag\{s_1,s_2,...,s_n\}$. Then the derivative of $V(e)$ with respect to time along the trajectory of system (5) is

$$\dot{V}(e)=e^T[(A(t)-K(t)+M(x,y))^T S+S(A(t)-K(t)+M(x,y))]\,e+2\Delta m^T Se. \quad (6)$$

Obviously,

$$\|\Delta m\|=\sqrt{\sum_{l=1}^{n}(f_l\sin(\omega_l t+\varphi_l)-\tilde{f}_l\sin(\tilde{\omega}_l t+\tilde{\varphi}_l))^2}$$

$$\le\sum_{l=1}^{n}\left|f_l\sin(\omega_l t+\varphi_l)-\tilde{f}_l\sin(\tilde{\omega}_l t+\tilde{\varphi}_l)\right|\le\sum_{l=1}^{n}(f_l+\tilde{f}_l)=\sum_{l=1}^{n}\tilde{f}_l\frac{f_l+\tilde{f}_l}{\tilde{f}_l}.$$

Letting $G_l=\dfrac{f_l+\tilde{f}_l}{\tilde{f}_l}$, we have

$$\|\Delta m\|\le\sum_{l=1}^{n}\tilde{f}_l G_l.$$

Hence

$$2\Delta m^T Se\le 2\|\Delta m^T Se\|\le 2\lambda_{max}\|\Delta m^T\|\|e\|\le 2\sum_{l=1}^{n}\tilde{f}_l G_l\lambda_{max}\|e\|,$$

where $\lambda_{\max}=\max\{s_i : i=1,2,\cdots n\}$ is the maximal eigenvalue of matrix $S$. Then system (6) can be rewritten as

$$\dot{V}=e^T[(A(t)-K(t)+M(x,y))^T S+S(A(t)-K(t)+M(x,y))+2\sum_{l=1}^{n}\tilde{f}_l\alpha I_n]e-2\sum_{l=1}^{n}\tilde{f}_l\alpha e^T e+2\Delta m^T Se,$$

where $\alpha\in R^+$.

Note that if

$$\|e\|>\frac{G\lambda_{\max}}{\alpha}\quad\text{with}\quad G=\max\{G_1,G_2,\cdots,G_n\}, \quad (7)$$

then

$$-2\sum_{l=1}^{n}\tilde{f}_l\alpha e^T e+2\Delta m^T Se\le 2\|e\|\sum_{l=1}^{n}\tilde{f}_l(G_l\lambda_{\max}-\alpha\|e\|)<0.$$

Furthermore, we assume that the coupling matrix $K(t)$ can be chosen such that

$$Q(t) = (A(t) - K(t) + M(x,y))^T S + S(A(t) - K(t) + M(x,y)) + \sum_{l=1}^{n} \tilde{f}_l \alpha I_n < 2, \quad (8)$$

which will be discussed in Section 4 in detail. Under the assumptions (7) and (8), we know that $\dot{V} < 0$. Suppose that the balls $E(a_0) = \{e : e^T e \le a_0\}$ with $a_0 = \frac{G^2 \lambda_{\max}^2}{\alpha^2}$ and $E(a_2) = \{e : e^T e \le a_2\}$ are respectively the smallest upper bound and the infimum of the ellipsoid $E(a_1) = \{e : e^T Se \le a_1\}$. Then we have

$$E(a_2) = \{e : e^T e \le a_2\} \supset E(a_1) = \{e : e^T Se \le a_1\} \supset E(a_0) = \{e : e^T e \le a_0\}.$$

The relationships among $E(a_0)$, $E(a_1)$ and $E(a_2)$ for a two-dimensional system are sketched in Fig.1. $\dot{V} < 0$ implies that the trajectory of the error system (5) will uniformly enter the ellipsoid $E(a_1)$ for any trajectory outside the ellipsoid. Hence the synchronization scheme (1)-(3) achieves synchronization with an error bound, denoted by $h$, if inequality (8) holds. The radius $\sqrt{a_2}$ of ball $E(a_2)$ is an enlarged synchronization error bound, namely, $h < \sqrt{a_2}$. Thus $H = \sqrt{a_2}$ is called the estimated synchronization error bound, which can be found out in the following.

Note that $S = diag\{s_1, s_2, \ldots, s_n\}$, then

$$e^T Se = s_1 e_1^2 + s_2 e_2^2 + \cdots + s_n e_n^2.$$

Let $\lambda_{\max}$ and $\lambda_{\min}$ be the maximal and minimal eigenvalue of matrix $S$ respectively. Hence, the maximal semiaxis $R_{\max}$ and minimal semiaxis $R_{\min}$ of the ellipsoid $E(a_1)$ are

$$R_{\max} = \frac{\sqrt{a_1}}{\sqrt{\lambda_{\min}}} = \sqrt{a_2}, \quad R_{\min} = \frac{\sqrt{a_1}}{\sqrt{\lambda_{\max}}} = \sqrt{a_0},$$

and then

$$H = \sqrt{a_2} = \frac{G\lambda_{\max}}{\alpha} \sqrt{\frac{\lambda_{\max}}{\lambda_{\min}}}.$$

From the above expression we know that $H$ reaches its minimum if $s_1 = s_2 = ... = s_n = s$ . In this situation, $\lambda_{\min} = \lambda_{\max} = s$ and the estimated synchronization error bound $H$ can be rewritten as

$$H = G\sigma,\ \sigma = \frac{s}{\alpha}. \tag{9}$$

Now we discuss some special cases of expression (9). If there exist only amplitude mismatch and phase mismatch between the external harmonic excitation of master system and that of slave one, i.e., $\tilde{\omega}_l = \omega_l$ , $l = 1,2,...,n$ , then

$$f_l \sin(\omega_l t + \varphi_l) - \tilde{f}_l \sin(\omega_l t + \tilde{\varphi}_l)$$
$$= (f_l - \tilde{f}_l)\sin(\omega_l t + \varphi_l) + \tilde{f}_l[\sin(\omega_l t + \varphi_l) - \sin(\omega_l t + \tilde{\varphi}_l)].$$

By the differential mean-value theorem, we have

$$\left| f_l \sin(\omega_l t + \varphi_l) - \tilde{f}_l \sin(\omega_l t + \tilde{\varphi}_l) \right|$$
$$\le \left| f_l - \tilde{f}_l \right| + \tilde{f}_l \left| \varphi_l - \tilde{\varphi}_l \right| = \tilde{f}_l \left( \frac{\left| f_l - \tilde{f}_l \right|}{\tilde{f}_l} + \left| \varphi_l - \tilde{\varphi}_l \right| \right).$$

As a result, we can replace $G_l$ in inequality (7) with

$$G_l = \frac{\left| f_l - \tilde{f}_l \right|}{\tilde{f}_l} + \left| \varphi_l - \tilde{\varphi}_l \right|. \tag{10}$$

Thus

$$G = \max\{G_1, G_2, \cdots, G_n\}$$
$$= \max\{\frac{\left| f_1 - \tilde{f}_1 \right|}{\tilde{f}_1} + \left| \varphi_1 - \tilde{\varphi}_1 \right|, \frac{\left| f_2 - \tilde{f}_2 \right|}{\tilde{f}_2} + \left| \varphi_2 - \tilde{\varphi}_2 \right|, \cdots, \frac{\left| f_n - \tilde{f}_n \right|}{\tilde{f}_n} + \left| \varphi_n - \tilde{\varphi}_n \right|\}.$$

Especially, if $\tilde{\varphi}_l = \varphi_l$, then

$$G = \max\{\frac{\left| f_1 - \tilde{f}_1 \right|}{\tilde{f}_1}, \frac{\left| f_2 - \tilde{f}_2 \right|}{\tilde{f}_2}, \cdots, \frac{\left| f_n - \tilde{f}_n \right|}{\tilde{f}_n}\}, \tag{11}$$

and if $\tilde{f}_l = f_l$, then

$$G = \max\{\left| \varphi_1 - \tilde{\varphi}_1 \right|, \left| \varphi_2 - \tilde{\varphi}_2 \right|, \cdots, \left| \varphi_n - \tilde{\varphi}_n \right|\}. \tag{12}$$

From Eqs.(9)-(12) we know that the synchronization error bound will be sufficiently small if the parameter mismatch turns to zero.

**4. Algebraic synchronization criteria**

Now we will deduce the condition under which Eq.(8) holds, that is, the matrix $Q(t)$ is negative definite. Assuming that

$$(A(t)+M(x,y))^T S+S(A(t)+M(x,y))=(sb_{ij}(t,x,y))_{n\times n} ,$$

where $b_{ij}(t,x,y)=b_{ji}(t,x,y)$, $i,j=1,2,\ldots,n$, then the matrix $Q(t)$ mentioned in expression (8) becomes

$$Q(t)=\begin{pmatrix} sb_{11}(t,x,y)-2k_1su_1+2\sum_{l=1}^{n}\tilde{f}_l\alpha & sb_{12}(t,x,y) & \cdots & sb_{1n}(t,x,y) \\ sb_{21}(t,x,y) & sb_{22}(t,x,y)-2k_2su_2+2\sum_{l=1}^{n}\tilde{f}_l\alpha & \cdots & sb_{2n}(t,x,y) \\ \cdots & \cdots & \cdots & \cdots \\ sb_{n1}(t,x,y) & sb_{n2}(t,x,y) & \cdots & sb_{nn}(t,x,y)-2k_nsu_n+2\sum_{l=1}^{n}\tilde{f}_l\alpha \end{pmatrix}.$$

Our object is to choose coupling coefficients $k_i$ such that the matrix $Q(t)$ is negative definite. To this end, we first introduce the Gerschgorin disc theorem.

**Lemma 1.** (Section 6.1 in Ref.49) Let $Z=(z_{ij})\in R^{n\times n}$ and $p_1,p_2,\cdots,p_n$ be positive numbers. Then all the eigenvalues of $Z$ lie in the region

$$\bigcup_{i=1}^{n}\{\lambda\in C:|\lambda-z_{ii}|\le\frac{1}{p_i}\sum_{\substack{j=1\\j\ne i}}^{n}p_j\left|z_{ij}\right|\},$$

where $C$ is the set of complex numbers.

According to Lemma 1, the matrix $Q(t)$ is negative definite if $k_i, p_i, \alpha$ and $s$ satisfy

$$2k_isu_i>sb_{ii}(t,x,y)+2\sum_{l=1}^{n}\tilde{f}_l\alpha+\frac{s}{p_i}\sum_{\substack{j=1\\j\ne i}}^{n}p_j\left|b_{ij}(t,x,y)\right|,\quad i=1,2,\ldots,n, \qquad (13)$$

namely

$$2k_iu_i>b_{ii}(t,x,y)+\frac{2G\sum_{l=1}^{n}\tilde{f}_l}{H}+\frac{1}{p_i}\sum_{\substack{j=1\\j\ne i}}^{n}p_j\left|b_{ij}(t,x,y)\right|,\quad i=1,2,\ldots,n. \qquad (14)$$

According to Definition 1, if the slave system is synchronized with the master with error bound $h$ ( $h\le H$ ), then $x_i-H\le y_i\le x_i+H, i=1,2,\cdots,n$ . Recall that the elements of $A(t)$ and $M(x,y)$ are bounded, there exist constants $b_{ij} > 0$ such that $\left|b_{ij}(t,x,y)\right| \le b_{ij}$, $i \ne j$, and constants $b_{ii}$ such that $b_{ii}(t,x,y) \le b_{ii}$, $i = 1,2,...,n$, for a prescribed error bound $H$. Substituting the bounds of $b_{ij}(t,x,y)$ and $b_{ii}(t,x,y)$ into the inequalities (14), we obtain more rigorous inequalities as **Hypothesis 2**

$$2k_i u_i > b_{ii} + \frac{2G\sum_{l=1}^{n}\tilde{f}_l}{H} + \frac{1}{p_i}\sum_{\substack{j=1 \\ j\ne i}}^{n} p_j b_{ij}, \quad i = 1,2,\ldots,n. \tag{15}$$

If the coupling coefficients $k_i$ and parameters $p_i$ are chosen such that inequalities (15) hold, then the matrix $Q(t)$ is negative definite, which implies that the trajectory of the error system (5) will uniformly enter the ellipsoid $E(a_1)$ (so the error bound $h \le H$) for any trajectory outside of the ellipsoid $E(a_1)$. We have therefore outlined a proof of the following theorem.

**Theorem 1** The synchronization scheme (1)-(2) achieves synchronization with error bound $h \le H$ if the **Hypotheses 1** and **2** are satisfied.

In the following, we will derive some easily verified synchronization conditions from Theorem 1. Note that $u_i(t) > 0$, $i = 1,2,\cdots,n$, if $(x_1, x_2, \cdots, x_n)$ and $(y_1, y_2, \cdots, y_n)$ are limited in the region $D = \{\left|x_i - y_i\right| < \pi, i = 1,2,\cdots,n\}$. Under this restriction, we can rewrite inequalities (15) as

$$k_i > \frac{1}{2u_i}\left(b_{ii} + \frac{2G\sum_{l=1}^{n}\tilde{f}_l}{H} + \frac{1}{p_i}\sum_{\substack{j=1 \\ j\ne i}}^{n} p_j b_{ij}\right), \quad i = 1,2,\ldots,n. \tag{16}$$

If the variables are further restricted in the region $D_0 = \{\left|x_i - y_i\right| \le \frac{3\pi}{4}, i = 1,2,\cdots,n\}$, then we have $1 \ge u_i(t) \ge \frac{2\sqrt{2}}{3\pi}$, $i = 1,2,\cdots,n$. As a result, simpler criteria can be obtained from inequalities (16) as follows:

$$k_i > \frac{3\pi}{4\sqrt{2}}(b_{ii} + \frac{2G\sum_{l=1}^{n}\tilde{f}_l}{H} + \frac{1}{p_i}\sum_{\substack{j=1\\ j\neq i}}^{n} p_j b_{ij}), \quad i = 1, 2, \ldots, n\,. \tag{17}$$

We will apply the criterion (17) instead of (15) or (16) in the upcoming example.

## 5. Analysis of synchronization error bound

In this section we will discover some relationships among the estimated synchronization error bound, coupling coefficient and parameter mismatch. Assume that there exists an $i_0 \in \{1, 2, \cdots, n\}$ such that

$$\max_{1\le i\le n}\{\frac{3\pi}{4\sqrt{2}}(b_{ii} + \frac{2G\sum_{l=1}^{n}\tilde{f}_l}{H} + \frac{1}{p_i}\sum_{\substack{j=1\\ j\neq i}}^{n} p_j b_{ij})\} = \frac{3\pi}{4\sqrt{2}}(b_{i_0 i_0} + \frac{2G\sum_{l=1}^{n}\tilde{f}_l}{H} + \frac{1}{p_{i_0}}\sum_{\substack{j=1\\ j\neq i_0}}^{n} p_j b_{i_0 j})\,.$$

For simplicity of notation, in what follows we denote $i_0$ as $i$. If the coupling coefficients $k_i$ are chosen identically, i.e., $K(t) = diag\{ku_1(t), ku_2(t), \cdots, ku_n(t)\}$, then the synchronization criteria (17) become

$$k > \frac{3\pi}{4\sqrt{2}}(b_{ii} + \frac{2G\sum_{l=1}^{n}\tilde{f}_l}{H} + \frac{1}{p_i}\sum_{\substack{j=1\\ j\neq i}}^{n} p_j b_{ij}), \quad i \in \{1, 2, ..., n\}\,. \tag{18}$$

In practice, we can take

$$k = \frac{3\pi}{4\sqrt{2}}(b_{ii} + \frac{2G\sum_{l=1}^{n}\tilde{f}_l}{H} + \frac{1}{p_i}\sum_{\substack{j=1\\ j\neq i}}^{n} p_j b_{ij}) + \varepsilon\,, \tag{19}$$

where $\varepsilon$ is a small positive constant. For the case of only amplitude mismatch in the external harmonic excitations, substituting expression (11) into the above inequalities yields

$$k = \frac{3\pi}{4\sqrt{2}}(b_{ii} + \frac{2\sum_{l=1}^{n}\tilde{f}_l\left|f - \tilde{f}\right|}{\tilde{f}\,H} + \frac{1}{p_i}\sum_{\substack{j=1\\ j\neq i}}^{n} p_j b_{ij}) + \varepsilon\,. \tag{20}$$

Hence we obtain the relationships among the estimated synchronization error bound $H$, the amplitude mismatch $\left|f - \tilde{f}\right|$ and the coupling coefficient $k$,

$$H=\frac{6\pi\left|f-\tilde{f}\right|\sum_{l=1}^{n}\tilde{f}_l}{\tilde{f}[4\sqrt{2}(k-\varepsilon)-3\pi b_{ii}-\frac{3\pi}{p_i}\sum_{\substack{j=1\\ j\neq i}}^{n}p_j b_{ij}]}. \tag{21}$$

For a fixed $k$, $H\to 0$ as $\left|f-\tilde{f}\right|\to 0$.

Similarly, for the case of only phase mismatch in the external harmonic excitations, substitution of expression (12) into expression (19) yields

$$k=\frac{3\pi}{4\sqrt{2}}(b_{ii}+\frac{2\left|\varphi-\tilde{\varphi}\right|\sum_{l=1}^{n}\tilde{f}_l}{H}+\frac{1}{p_i}\sum_{\substack{j=1\\ j\neq i}}^{n}p_j b_{ij})+\varepsilon\,. \tag{22}$$

Hence we obtain the relationships among the estimated synchronization error bound $H$, the phase mismatch $\left|\varphi-\tilde{\varphi}\right|$ and the coupling coefficient $k$,

$$H=\frac{6\pi\left|\varphi-\tilde{\varphi}\right|\sum_{l=1}^{n}\tilde{f}_l}{4\sqrt{2}(k-\varepsilon)-3\pi b_{ii}-\frac{3\pi}{p_i}\sum_{\substack{j=1\\ j\neq i}}^{n}p_j b_{ij}}\,. \tag{23}$$

For a fixed $k$, $H\to 0$ as $\left|\varphi-\tilde{\varphi}\right|\to 0$.

## 6. Numerical example

Before starting with the example, we first outline the design procedure as follows:

1) Construct the slave system associated with the master system via sinusoidal state error feedback control.
2) Verify **Hypothesis 1** to be satisfied.
3) For a prescribe error bound $H$, determine the inequalities (17) instead of **Hypothesis 2** to ensure that the synchronization scheme (1)-(3) achieves synchronization with error bound $h\leq H$.
4) Analyse the relations among the estimated error bound, coupling coefficient and parameter mismatch.
5) Verify the control robustness against particular parameter mismatches.

Consider a gyrostat system[22]

$$\begin{aligned}\dot{x}_1 &= -x_2x_3 - 0.5(1+6.5\cos t)x_2 + 0.4x_3 - 0.002(x_1 + x_1^3) \\ \dot{x}_2 &= x_1x_3 - 0.4x_3 + 0.5(1+6.5\cos t)x_1 - 0.002(x_2 + x_2^3) \\ \dot{x}_3 &= -0.2x_1 + 0.2x_2 - 0.2x_3 - 0.002(x_3 + x_3^3) + 1.625\sin t\end{aligned} \quad . \tag{24}$$

Obviously, $m(t)$ is $(0,0,m_3(t))^T$ with $m_3(t) = f_3\sin(\omega_3 t + \varphi_3) = 1.625\sin t$, compared with system (1). The gyrostat system exhibits chaos behavior, as shown in Fig. 2(refer to Ref.22 for more details). Subjected to a sinusoidal error feedback control, the slave system is constructed as follows:

$$\begin{aligned}\dot{y}_1 &= -y_2y_3 - 0.5(1+6.5\cos t)y_2 + 0.4y_3 - 0.002(y_1 + y_1^3) + k_1\sin(x_1 - y_1) \\ \dot{y}_2 &= y_1y_3 - 0.4y_3 + 0.5(1+6.5\cos t)y_1 - 0.002(y_2 + y_2^3) + k_2\sin(x_2 - y_2) \\ \dot{y}_3 &= -0.2y_1 + 0.2y_2 - 0.2y_3 - 0.002(y_3 + y_3^3) + \tilde{f}_3\sin(\tilde{\omega}_3 t + \tilde{\varphi}_3) + k_3\sin(x_3 - y_3)\end{aligned} \quad . \tag{25}$$

The error variable $e = (x_1 - y_1, x_2 - y_2, x_3 - y_3)^T$ satisfies

$$\dot{e} = (A(t) - K(t) + M(x,y))e + \Delta m$$

with

$$A(t) = \begin{pmatrix} -0.002 & -0.5(1+6.5\cos t) & 0.4 \\ 0.5(1+6.5\cos t) & -0.002 & -0.4 \\ -0.2 & 0.2 & -0.202 \end{pmatrix}, \quad K(t) = \begin{pmatrix} k_1u_1 & 0 & 0 \\ 0 & k_2u_2 & 0 \\ 0 & 0 & k_3u_3 \end{pmatrix},$$

$$u_i(t) = \frac{\sin(x_i - y_i)}{x_i - y_i}, \quad i = 1,2,3,$$

$$M(x,y) = \begin{pmatrix} -0.002(x_1^2 + x_1y_1 + y_1^2) & -y_3 & -x_2 \\ y_3 & -0.002(x_2^2 + x_2y_2 + y_2^2) & x_1 \\ 0 & 0 & -0.002(x_3^2 + x_3y_3 + y_3^2) \end{pmatrix}.$$

Hence

$$Q(t) = (A(t) - K(t) + M(x,y))^T S + S(A(t) - K(t) + M(x,y)) + 2\tilde{f}_3\alpha I_2$$

$$= \begin{pmatrix} \begin{matrix} s(-0.004 - 0.004(x_1^2 + x_1y_1 + y_1^2)) \\ -2sk_1u_1 + 2\tilde{f}_3\alpha \end{matrix} & 0 & s(0.2 - x_2) \\ 0 & \begin{matrix} s(-0.004 - 0.004(x_2^2 + x_2y_2 + y_2^2)) \\ -2sk_2u_2 + 2\tilde{f}_3\alpha \end{matrix} & s(-0.2 + x_1) \\ s(0.2 - x_2) & s(-0.2 + x_1) & \begin{matrix} s(-0.404 - 0.004(x_3^2 + x_3y_3 + y_3^2)) \\ -2sk_3u_3 + 2\tilde{f}_3\alpha \end{matrix} \end{pmatrix}.$$

Under the assumption that the slave system can be synchronized with the master one with error bound $h$ ($h \le H$), there exists a $T_0 > 0$ such that $|x_i - y_i| < H, i = 1,2,3$, for $t > T_0$. Hence,

$$b_{11}(t,x,y) = -0.004 - 0.004(x_1^2 + x_1 y_1 + y_1^2)$$

$$= -0.004 x_1^2 - 0.004 x_1 y_1 - 0.004 y_1^2 - 0.004$$

$$\le -0.004 x_1 (x_1 \pm H) - 0.004$$

$$= -0.004 (x_1 \pm \frac{H}{2})^2 + 0.001 H^2 - 0.004$$

$$\le 0.001 H^2 - 0.004 .$$

The estimated synchronization error bound $H$ can be chosen according to the demand in practical application. In this paper, it is set to be $0 < H \le 1$, then we have

$$b_{11}(t,x,y) = -0.004 - 0.004(x_1^2 + x_1 y_1 + y_1^2) \le 0.001 - 0.004 = -0.003 .$$

Similarly,

$$b_{22}(t,x,y) = -0.004 - 0.004(x_2^2 + x_2 y_2 + y_2^2) \le 0.001 - 0.004 = -0.003 ,$$

$$b_{33}(t,x,y) = -0.404 - 0.004(x_3^2 + x_3 y_3 + y_3^2) \le 0.001 - 0.404 = -0.403 .$$

Therefore, the synchronization criteria corresponding to inequalities (17) can be obtained as follows:

$$k_1 > \frac{3\pi}{4\sqrt{2}}(-0.003 + \frac{2\tilde{f}_3}{H} G + \frac{p_3}{p_1}|0.2 - x_2|) ,$$

$$k_2 > \frac{3\pi}{4\sqrt{2}}(-0.003 + \frac{2\tilde{f}_3}{H} G + \frac{p_3}{p_2}|-0.2 + x_1|) , \tag{26}$$

$$k_3 > \frac{3\pi}{4\sqrt{2}}(-0.403 + \frac{2\tilde{f}_3}{H} G + \frac{p_1}{p_3}|0.2 - x_2| + \frac{p_2}{p_3}|-0.2 + x_1|) .$$

Thus we conclude that the synchronization scheme (24)-(25) can achieve synchronization with error bound $h \le H$ if the coupling coefficients $k_1, k_2$ and $k_3$ satisfy the inequalities (26).

From Fig.2, we know that the bounds of the chaotic attractor are $-3 < x_1 < 2.4$, $-3 < x_2 < 2$ and $-2 < x_3 < 1.4$. If we take $\tilde{f}_3 = 1.6$, $\tilde{\omega}_3 = 0.9$ and $\tilde{\varphi}_3 = 0.1$, namely, the amplitudes, frequencies and phases of external harmonic excitations of two coupled systems are mismatched simultaneously, then $G = \frac{|f_3 + \tilde{f}_3|}{\tilde{f}_3} = 2.02$. For a prescribed estimated synchronization error bound，such as $H = 0.1$, and $p_1 = p_2 = p_3$, we can get $k_1 > 113.288$， $k_2 > 113.288$ and $k_3 > 117.964$ from criteria (26). The simulation in Fig.3 shows that the synchronization scheme (24)-(25) achieves synchronization with error bound $h < H = 0.1$, where the coupling coefficients are $k_1 = 113.29$ ， $k_2 = 113.29$ and $k_3 = 117.97$ ， and the initial values are $(x_1(0), x_2(0), x_3(0)) = (1,1,1)$ and $(y_1(0), y_2(0), y_3(0)) = (-1,-1,-1)$.

If the frequencies of two coupled systems are same, we take $\tilde{f}_3 = 1.6$， $\tilde{\omega}_3 = \omega_3 = 1$ and $\tilde{\varphi}_3 = 0.1$, then we can get $G = \frac{|f_3 - \tilde{f}_3|}{\tilde{f}_3} + |\varphi_3 - \tilde{\varphi}_3| = 0.12$. If we choose $H = 0.01$ and $p_1 = p_2 = p_3$, we can get $k_1 > 69.467$, $k_2 > 69.467$ and $k_3 > 74.143$ from criteria (26). As shown in Fig.4, the synchronization scheme (24)-(25) can achieve synchronization with error bound $h < H = 0.01$, where the coupling coefficients are $k_1 = 69.47$， $k_2 = 69.47$ and $k_3 = 74.15$, and the initial conditions are the same as in Fig.3.

If there exists only amplitude mismatch between two coupled systems, we take $\tilde{f}_3 = 1.6$， $\tilde{\omega}_3 = \omega_3 = 1$ and $\tilde{\varphi}_3 = \varphi_3 = 0$, then $G = \frac{|f_3 - \tilde{f}_3|}{\tilde{f}_3} = 0.02$. If $H = 0.01$ and $p_1 = p_2 = p_3$， we can get $k_1 > 16.027$， $k_2 > 16.027$ and $k_3 > 20.703$ from criteria (26). Fig.5 shows that the synchronization scheme (24)-(25) achieves synchronization with error bound $h < H = 0.01$, where the coupling coefficients are $k_1 = 16.03$, $k_2 = 16.03$ and $k_3 = 20.71$, and the initial conditions are the same as in Fig.3.

In the case of $\tilde{\omega}_3 = \omega_3$ and $\tilde{\varphi}_3 = \varphi_3$, we know that $G = \frac{|f_3 - \tilde{f}_3|}{\tilde{f}_3}$ from expression (11). Taking $p_1 = p_2 = 100$ and $p_3 = 135$, another criteria for synchronizing systems (24)

and (25) can be obtained from inequalities (26)

$$k_1 > \frac{3\pi}{4\sqrt{2}}(-0.003 + \frac{2\left|f_3 - \tilde{f}_3\right|}{H} + \frac{135}{100}\left|0.2 - x_2\right|),$$

$$k_2 > \frac{3\pi}{4\sqrt{2}}(-0.003 + \frac{2\left|f_3 - \tilde{f}_3\right|}{H} + \frac{135}{100}\left|-0.2 + x_1\right|), \tag{27}$$

$$k_3 > \frac{3\pi}{4\sqrt{2}}(-0.403 + \frac{2\left|f_3 - \tilde{f}_3\right|}{H} + \frac{100}{135}\left|0.2 - x_2\right| + \frac{100}{135}\left|-0.2 + x_1\right|).$$

Substituting the bounds of the strange attractor of the gyrostat system into inequalities (27), we obtain more rigorous inequalities as follows

$$k_1 > 7.209 + \frac{3.34\left|f_3 - \tilde{f}_3\right|}{H},$$

$$k_2 > 7.209 + \frac{3.34\left|f_3 - \tilde{f}_3\right|}{H}, \tag{28}$$

$$k_3 > 7.244 + \frac{3.34\left|f_3 - \tilde{f}_3\right|}{H}.$$

The following criterion is corresponding to criteria (20)

$$k = k_3 = 7.244 + \frac{3.34\left|f_3 - \tilde{f}_3\right|}{H} + \varepsilon. \tag{29}$$

For simulation, $\varepsilon = 0.001$ is chosen. The estimated error bound $H$ corresponding to expression (21) becomes

$$H = \frac{3.34\left|f_3 - \tilde{f}_3\right|}{k - 7.245}. \tag{30}$$

Taking $\left|f_3 - \tilde{f}_3\right| = 0.02$ as an example，then

$$H = \frac{0.0668}{k - 7.245}. \tag{31}$$

For different coupling coefficients $k$, the estimated synchronization error bounds (31) are compared with numerical error bounds, which is shown in Fig.6. The numerical synchronization error bound is calculated by $\|e\|_{\max} = \max_{T_0 \le t \le T} \sqrt{(x_1 - y_1)^2 + (x_2 - y_2)^2 + (x_3 - y_3)^2}$, where $T_0$ represents a time threshold after which the synchronization error will be stabilized in a prescribed error bound and $T$ is a sufficiently large positive number. In the following simulations, $T_0 = 200$ and $T = 10000$ time unit are taken.

For physical meaning, we calculate the relationship between the relative error bound $H(\%) = \frac{H}{\|x\|_{\max}}$ and relative parameter mismatch $\Delta f(\%) = \frac{|f - \tilde{f}|}{f}$ as follows:

$$H(\%) = \frac{3.34 f_3}{\|x\|_{\max}(k - 7.245)} \Delta f_3(\%), \tag{32}$$

where $\|x\|_{\max}$ is defined as $\|x\|_{\max} = \max_{T_0 \le t \le T} \sqrt{x_1^{\ 2} + x_2^{\ 2} + x_3^{\ 2}}$. For the gyrostat system (24), $\|x\|_{\max} = 3.5$. Taking $k = 15$ as an example, then

$$H(\%) = 0.2 \Delta f_3(\%). \tag{33}$$

Let $\|e\|_{\max}(\%) = \frac{\|e\|_{\max}}{\|x\|_{\max}}$. The estimated synchronization error bounds $H(\%)$ are compared with the numerical error bounds $\|e\|_{\max}(\%)$ versus different amplitude mismatches $\Delta f_3(\%)$. The comparison result is shown in Fig.7, from which we know that a 12% amplitude mismatch induces approximately only a 2.4% synchronization error, which shows that the present control is robust against amplitude mismatch.

If there exists only phase mismatch between the two coupled systems, we take $\tilde{f}_3 = f_3 = 1.625$, $\tilde{\omega}_3 = \omega_3 = 1$ and $\tilde{\varphi}_3 = 0.1$. Hence $G = |\varphi_3 - \tilde{\varphi}_3| = 0.1$. We can get $k_1 > 59.614$, $k_2 > 59.614$, $k_3 > 64.29$ from criteria (26) if $H = 0.01$ and $p_1 = p_2 = p_3$. Fig.8 shows that the synchronization scheme (24)-(25) achieves synchronization with error bound $h < H = 0.01$. The coupling coefficients are $k_1 = 59.62$, $k_2 = 59.62$ and $k_3 = 64.3$, and the initial conditions are the same as in Fig.3.

For the case of $\tilde{f}_3 = f_3, \tilde{\omega}_3 = \omega_3$, from expression (12) we know that $G = |\varphi_3 - \tilde{\varphi}_3|$. If $p_1 = p_2 = 100$, $p_3 = 135$, another criteria for synchronizing the systems (24) and (25) can be obtained from inequalities (26)

$$k_1 > \frac{3\pi}{4\sqrt{2}}(-0.003 + \frac{2\tilde{f}_3|\varphi_3 - \tilde{\varphi}_3|}{H} + \frac{135}{100}|0.2 - x_2|)\,,$$

$$k_2 > \frac{3\pi}{4\sqrt{2}}(-0.003 + \frac{2\tilde{f}_3|\varphi_3 - \tilde{\varphi}_3|}{H} + \frac{135}{100}|-0.2 + x_1|)\,, \tag{34}$$

$$k_3 > \frac{3\pi}{4\sqrt{2}}(-0.403 + \frac{2\tilde{f}_3|\varphi_3 - \tilde{\varphi}_3|}{H} + \frac{100}{135}|0.2 - x_2| + \frac{100}{135}|-0.2 + x_1|)\,.$$

Substituting the bounds of the chaotic attractor of the gyrostat system into inequalities (34) leads to more rigorous inequalities

$$k_1 > 7.209 + \frac{5.428|\varphi_3 - \tilde{\varphi}_3|}{H}\,,$$

$$k_2 > 7.209 + \frac{5.428|\varphi_3 - \tilde{\varphi}_3|}{H}\,, \tag{35}$$

$$k_3 > 7.244 + \frac{5.428|\varphi_3 - \tilde{\varphi}_3|}{H}\,.$$

The following criterion is corresponding to criteria (22)

$$k = k_3 = 7.244 + \frac{5.428|\varphi_3 - \tilde{\varphi}_3|}{H} + \varepsilon\,. \tag{36}$$

In simulation, $\varepsilon = 0.001$ is chosen as above. The estimated error bound $H$ corresponding to expression (23) becomes

$$H = \frac{5.428|\varphi_3 - \tilde{\varphi}_3|}{k - 7.245}\,. \tag{37}$$

Taking $|\varphi_3 - \tilde{\varphi}_3| = 0.1$ as an example，then

$$H = \frac{0.5428}{k - 7.245}\,. \tag{38}$$

For different coupling coefficients $k$, the estimated synchronization error bounds (38) are compared with the numerical error bounds, shown in Fig.9.

Similar to the discussion of Eq.(32), letting $\Delta\varphi_3(\%) = \frac{|\varphi_3 - \tilde{\varphi}_3|}{\varphi_3}$, we can obtain the relative error bound as follows:

$$H(\%) = \frac{5.428\varphi_3}{\|x\|_{\max}(k-7.245)}\Delta\varphi_3(\%). \quad (39)$$

Taking $\varphi_3 = 2\pi$ and $k = 50$ as an example, then

$$H(\%) = 0.228\Delta\varphi_3(\%). \quad (40)$$

For different phase mismatches $\Delta\varphi_3(\%)$, the estimated synchronization error bounds (40) are compared with the numerical error bounds $\|e\|_{\max}(\%)$, shown in Fig.10. From the comparison we know that a 12% phase mismatch induces approximately a 2.7% synchronization error. So we conclude that the present control is robust against phase mismatch.

## 7. Conclusions

In this paper we have investigated in detail the synchronization of a class of master-slave non-autonomous chaotic systems coupled by a sinusoidal state error feedback controller, where parameter mismatch exists between the external harmonic excitation of master system and that of slave one. A concept of synchronization with error bound is introduced. The synchronization criteria are in the form of algebraic inequalities, so they are convenient in applications. The bounds of synchronization error are estimated analytically. It follows from the estimated synchronization error bound that the synchronization error can be controlled as small as possible by increasing the coupling strength or decreasing the magnitude of mismatch. The three-dimensional chaotic gyrostat system is used as an example to verify the effectiveness of the proposed synchronization criteria. The comparisons of estimated synchronization error bounds with the numerical error bounds show that a 12% amplitude mismatch or phase mismatch induces acceptable small synchronization errors, which verifies that the present control is robust against amplitude or phase mismatch.

**Acknowledgements** Research is supported by the National Natural Science

Foundation of China under grant No 60674049, and Foundation for supporting Universities in Fujian Province under grant No 2007F5099.

## Figure captions

Fig.1 Illustration of circles $E(a_0)$, $E(a_2)$ and ellipse $E(a_1)$

Fig.2 Phase portrait of the gyrostat system (24) showing the strange attractor

Fig.3 Error between the master-slave systems (24)-(25) with $k_1 = 113.29$, $k_2 = 113.29$ and $k_3 = 117.97$ when amplitude, frequency and phase are mismatched simultaneously. The beeline and curve represent the estimated error bound $H = 0.1$ and numerical error $\|e(t)\|$, respectively.

Fig.4 Error between the master-slave systems (24)-(25)with $k_1 = 69.47$, $k_2 = 69.47$ and $k_3 = 74.15$ when amplitude and phase are mismatched. The beeline and curve represent the estimated error bound $H = 0.01$ and numerical error $\|e(t)\|$, respectively.

Fig.5 Error between the master-slave systems (24)-(25) with $k_1 = 16.03$, $k_2 = 16.03$ and $k_3 = 20.71$ when there exists only amplitude mismatch. The beeline and curve represent the estimated error bound $H = 0.01$ and numerical error $\|e(t)\|$, respectively.

Fig.6 Synchronization error bounds versus the different coupling coefficients $k$. The solid curve and dotted curve represent the estimated synchronization error bound $H$ and the numerical error bound $\|e\|_{\max}$, respectively.

Fig.7 Synchronization error bounds versus the different amplitude mismatches $\Delta f_3(\%)$. The solid curve and dotted curve represent the estimated synchronization error bound $H(\%)$ and the numerical error bound $\|e\|_{\max}(\%)$, respectively.

Fig.8 Error between the master-slave systems (24)-(25) with $k_1 = 59.62$, $k_2 = 59.62$ and $k_3 = 64.3$ when there exists only phase mismatch. The beeline and curve represent the estimated error bound $H = 0.01$ and numerical error $\|e(t)\|$, respectively.

Fig.9 Synchronization error bounds versus the different coupling coefficients $k$. The solid curve and dotted curve represent the estimated synchronization error bound $H$

and the numerical error bound $\|e\|_{\max}$, respectively.

Fig.10 Synchronization error bounds versus the different phase mismatches $\Delta\varphi_3(\%)$.

The solid curve and dotted curve represent the estimated synchronization error bound $H(\%)$ and the numerical error bound $\| e\|_{\max}(\%)$, respectively.

# Figures

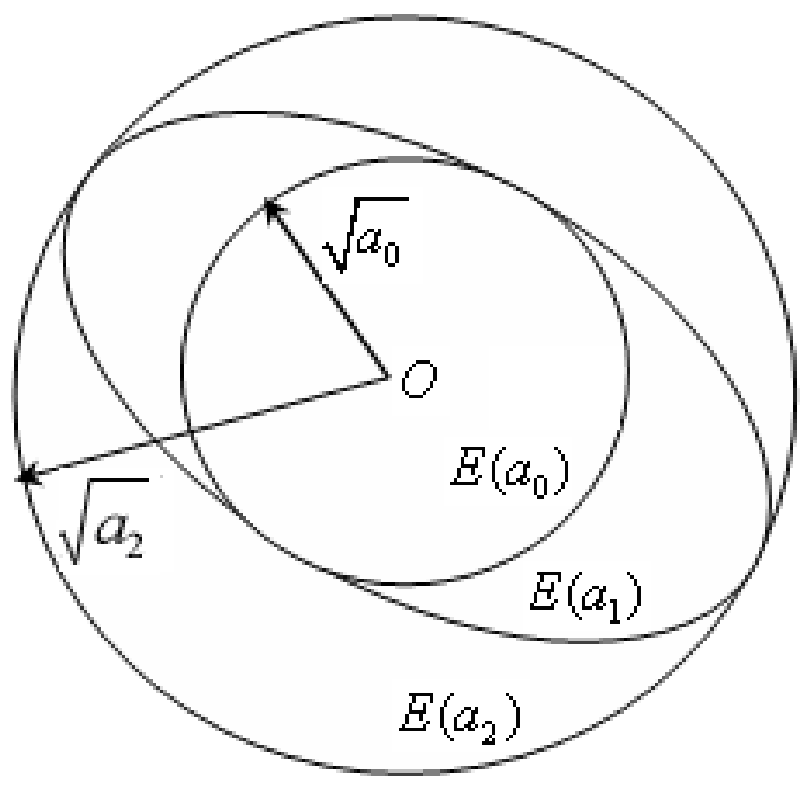

Fig.1 Illustration of circles $E(a_0)$ and $E(a_2)$, and ellipse $E(a_1)$

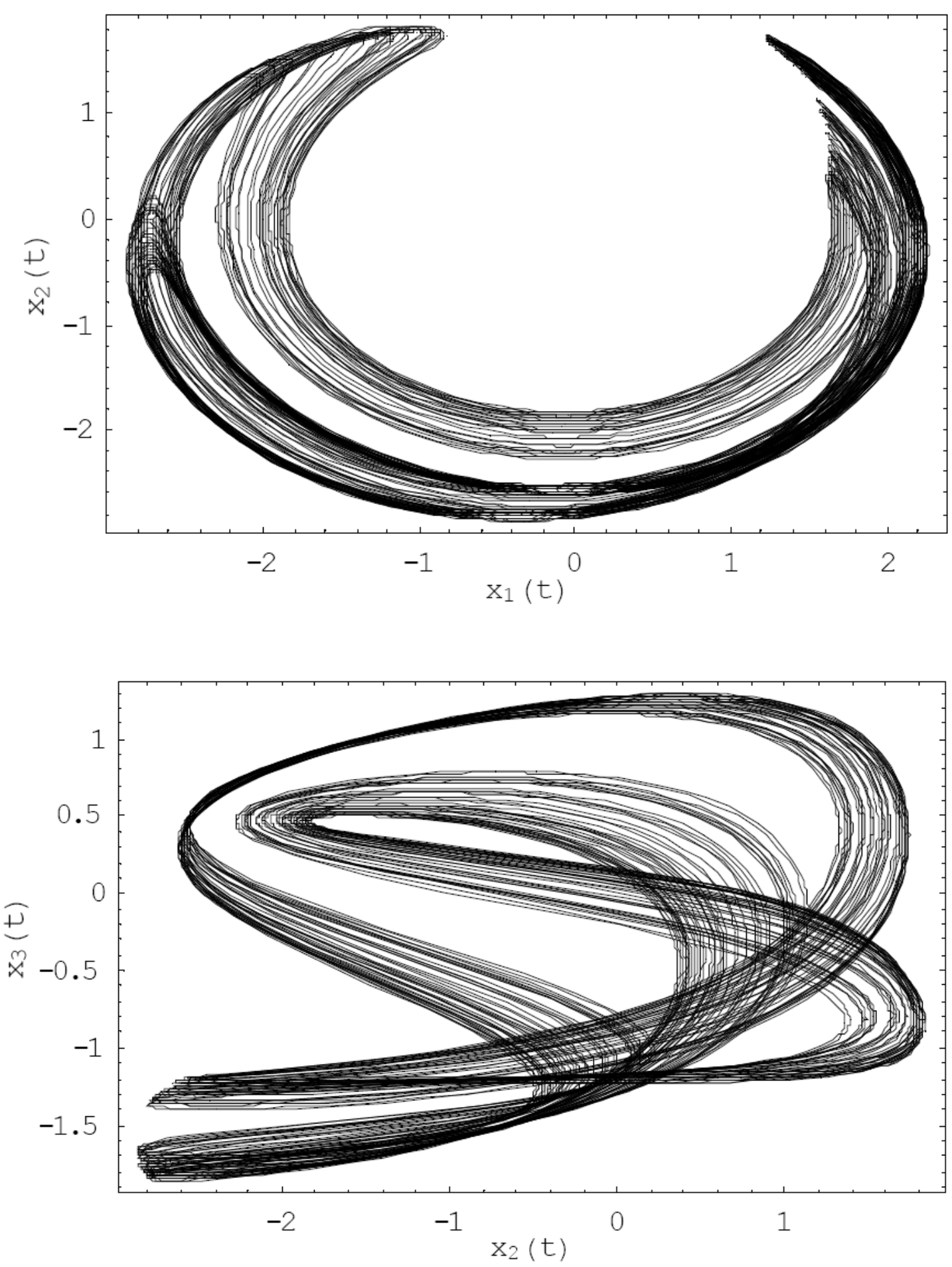

Fig.2 Phase portrait of the gyrostat system (24) showing the chaotic attractor

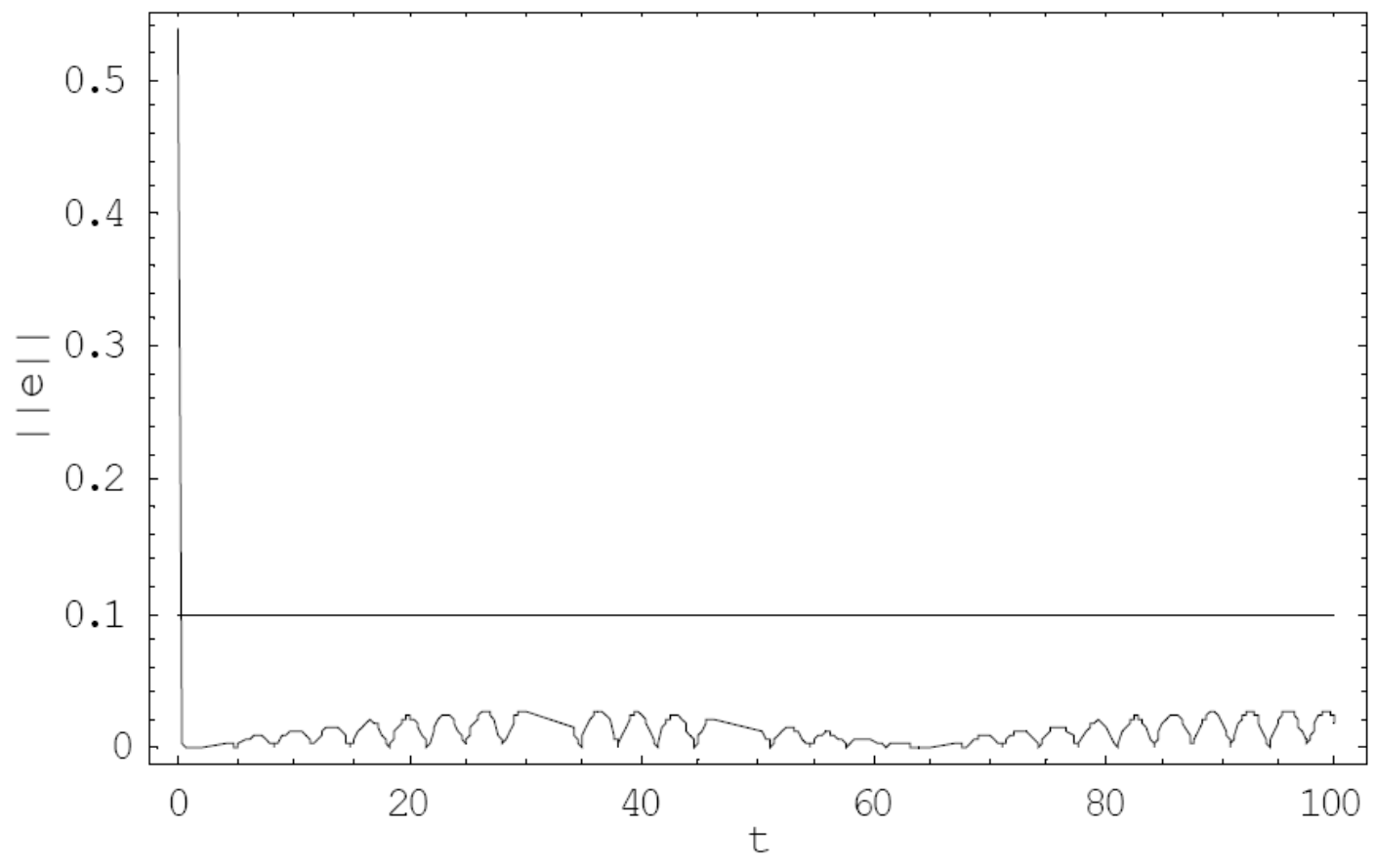

Fig.3 Error between the master-slave systems (24)-(25) with $k_1 = 113.29$ , $k_2 = 113.29$ and $k_3 = 117.97$ when amplitude, frequency and phase are mismatched simultaneously. The beeline and curve represent the estimated error bound $H = 0.1$ and numerical error $\|e(t)\|$, respectively.

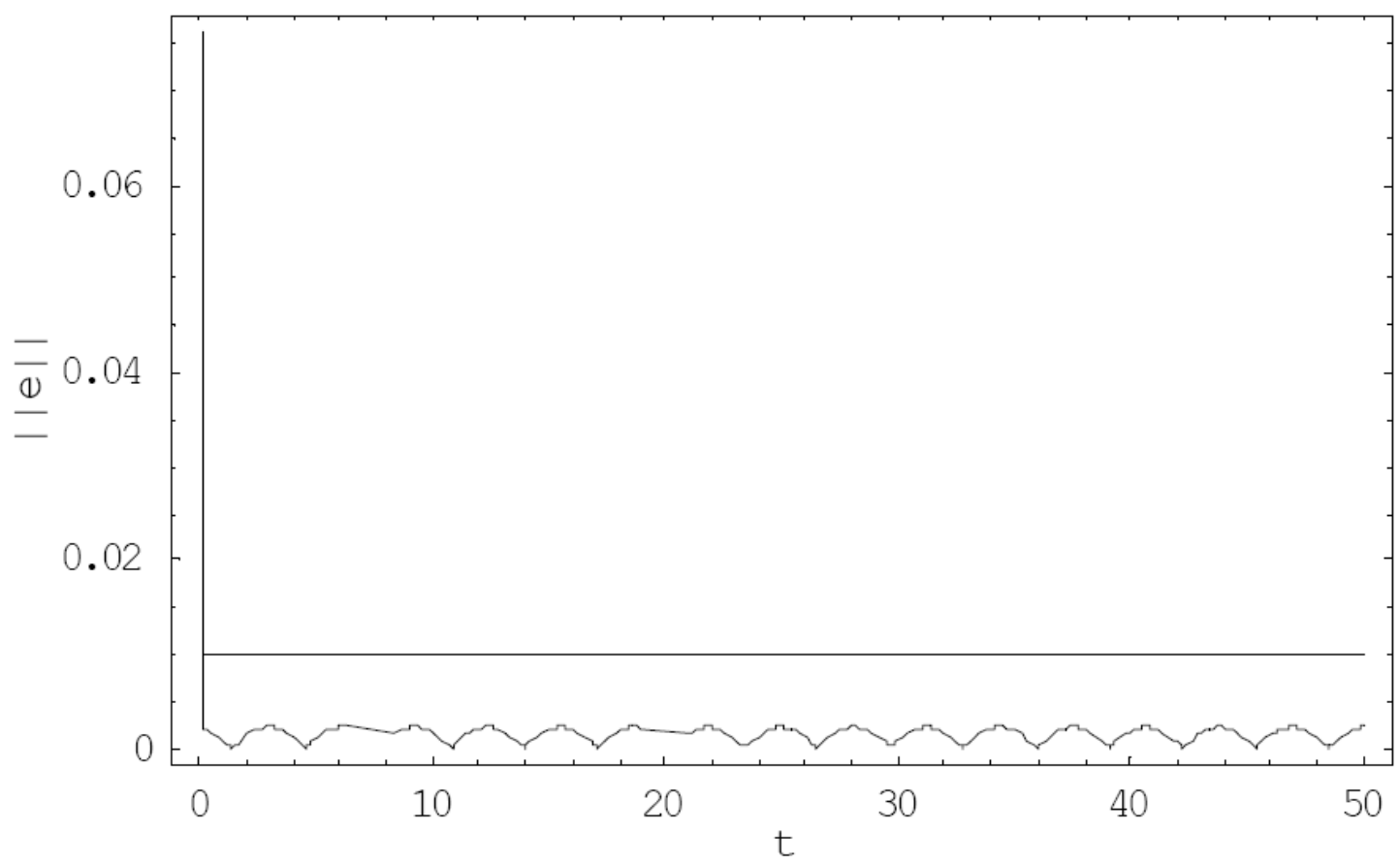

Fig.4 Error between the master-slave systems (24)-(25) with $k_1 = 69.47$ , $k_2 = 69.47$ and $k_3 = 74.15$ when amplitude and phase are mismatched. The beeline and curve represent the estimated error bound $H = 0.01$ and numerical error $\|e(t)\|$, respectively.

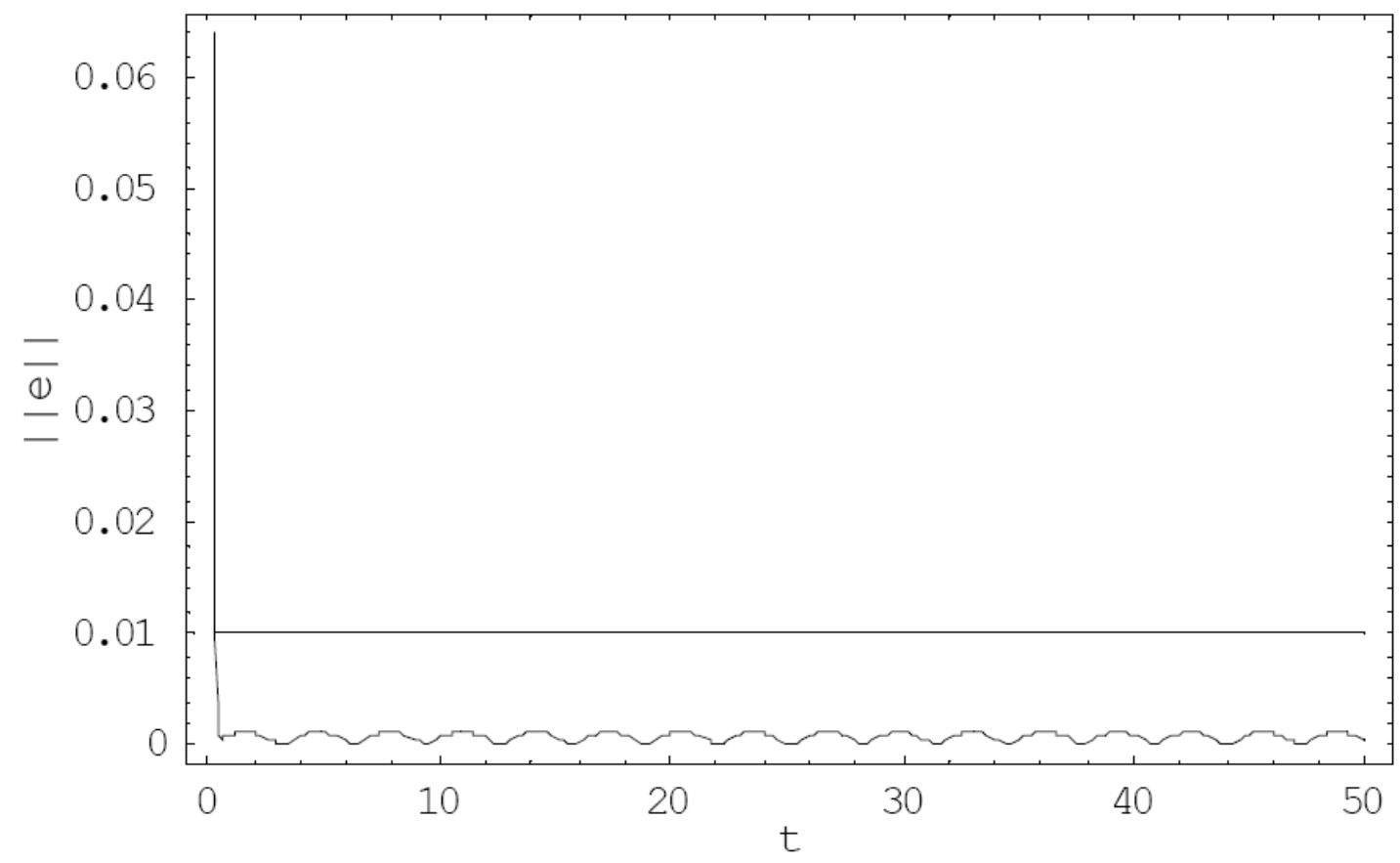

Fig.5 Error between the master-slave systems (24)-(25) with $k_1 = 16.03$, $k_2 = 16.03$ and $k_3 = 20.71$ when there exists only amplitude mismatch. The beeline and curve represent the estimated error bound $H = 0.01$ and numerical error $\|e(t)\|$, respectively.

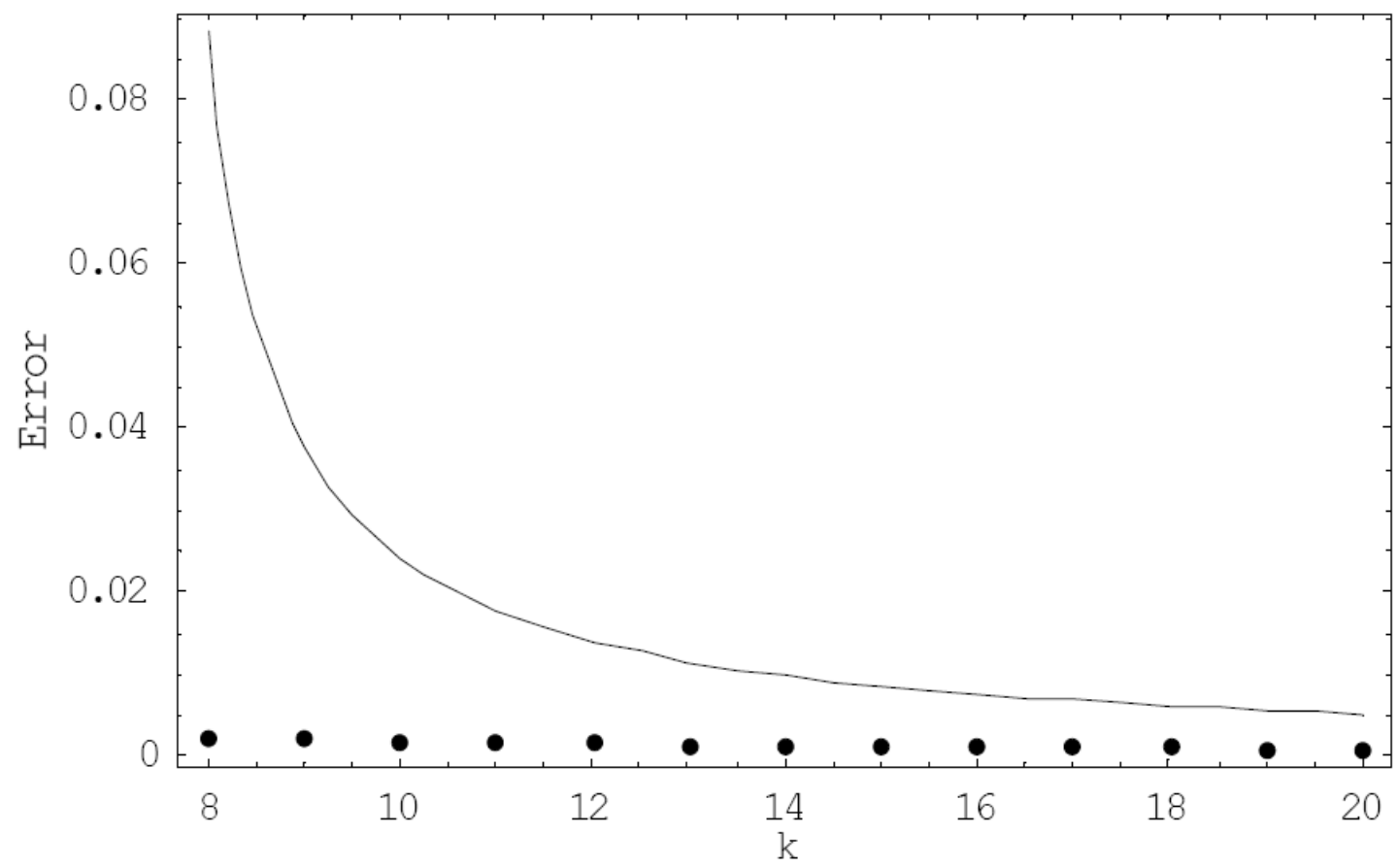

Fig.6 Synchronization error bounds versus the different coupling coefficients $k$. The solid curve and dotted curve represent the estimated synchronization error bound $H$ and the numerical error bound $\|e\|_{\max}$, respectively.

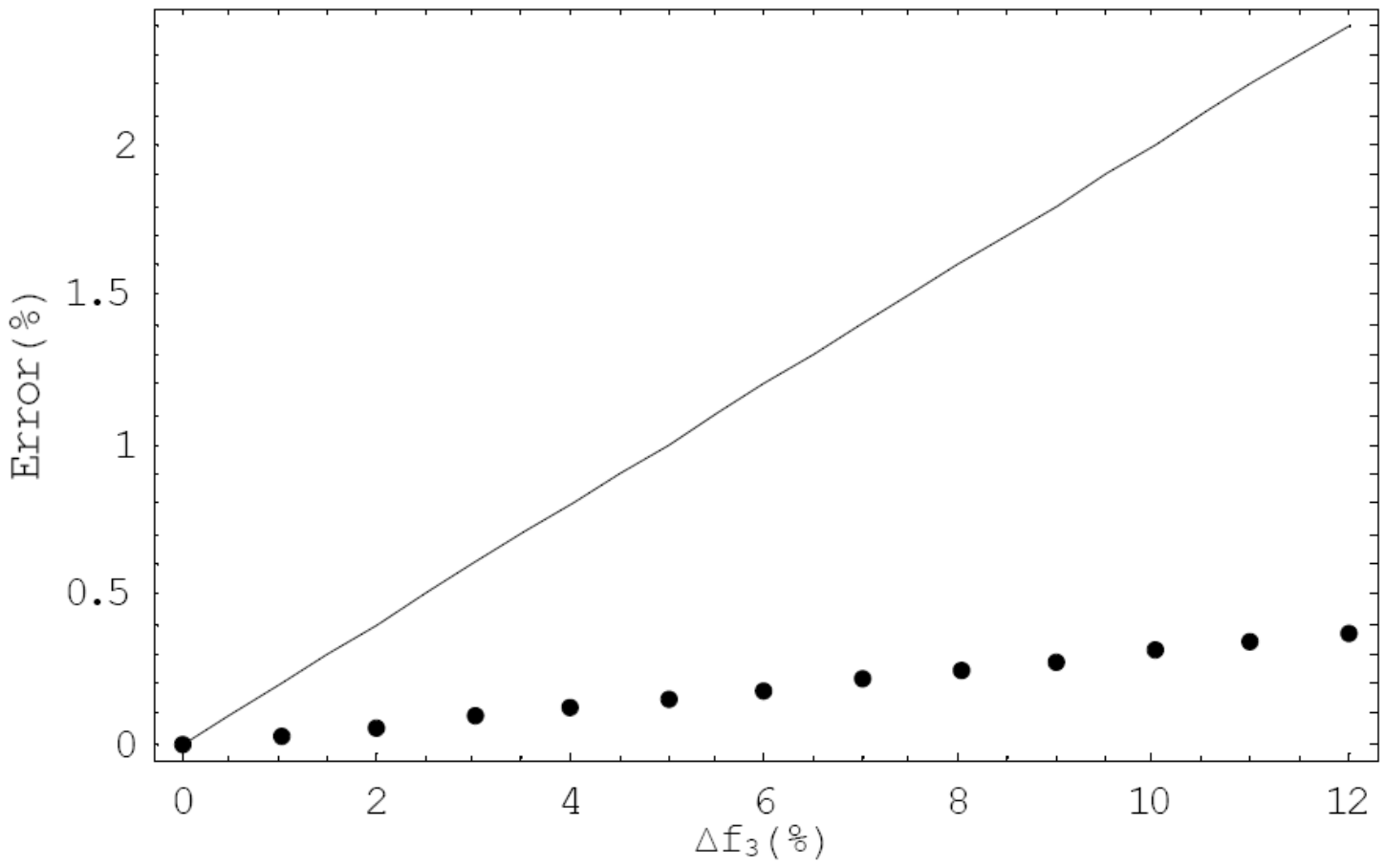

Fig.7 Synchronization error bounds versus the different amplitude mismatches $\Delta f_3(\%)$. The solid curve and dotted curve represent the estimated synchronization error bound $H(\%)$ and the numerical error bound $\|e\|_{\max}(\%)$, respectively.

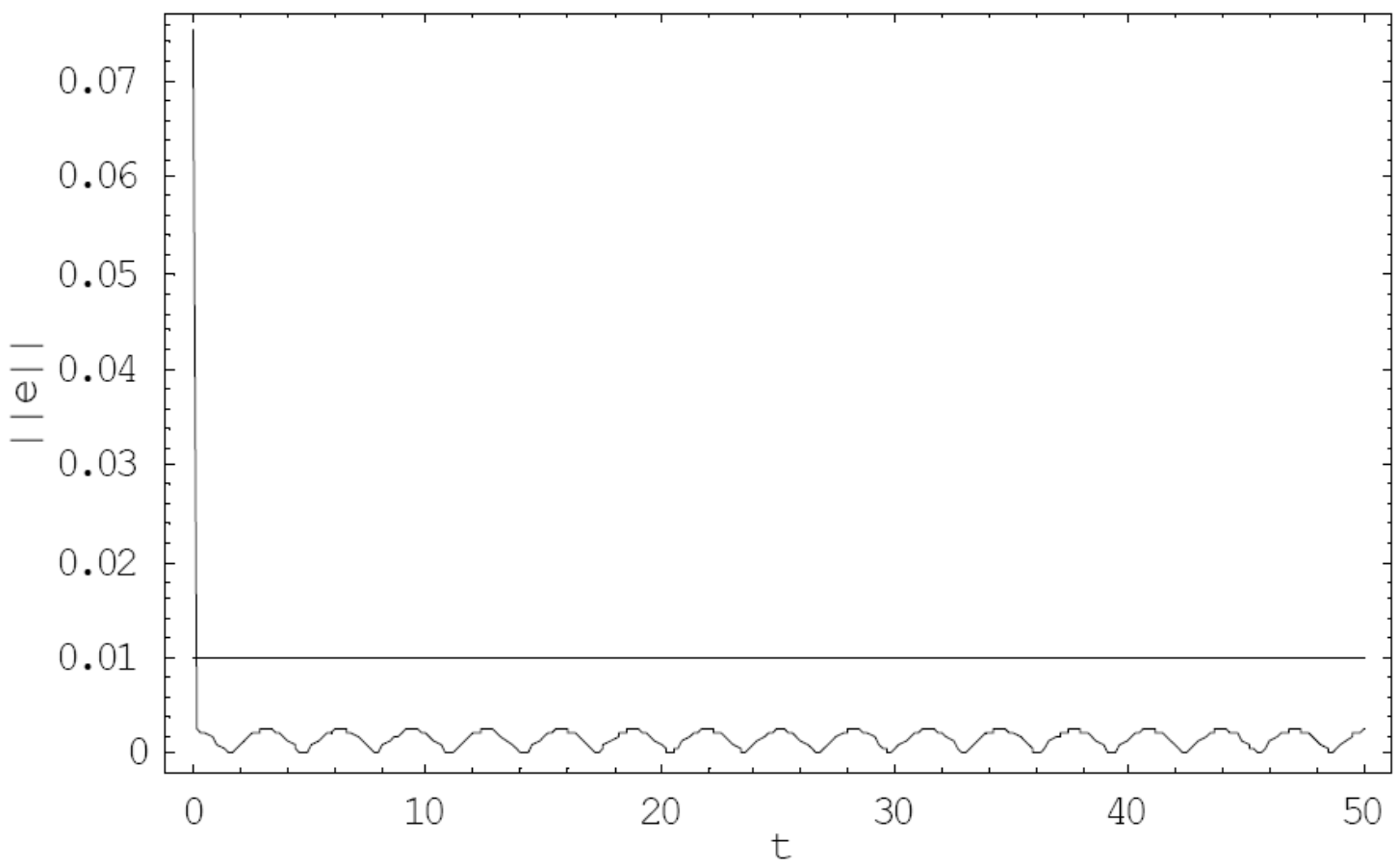

Fig.8 Error between the master-slave systems (24)-(25) with $k_1 = 59.62$, $k_2 = 59.62$ and $k_3 = 64.3$ when there exists only phase mismatch. The beeline and curve represent the estimated error bound $H = 0.01$ and numerical error $\|e(t)\|$, respectively.

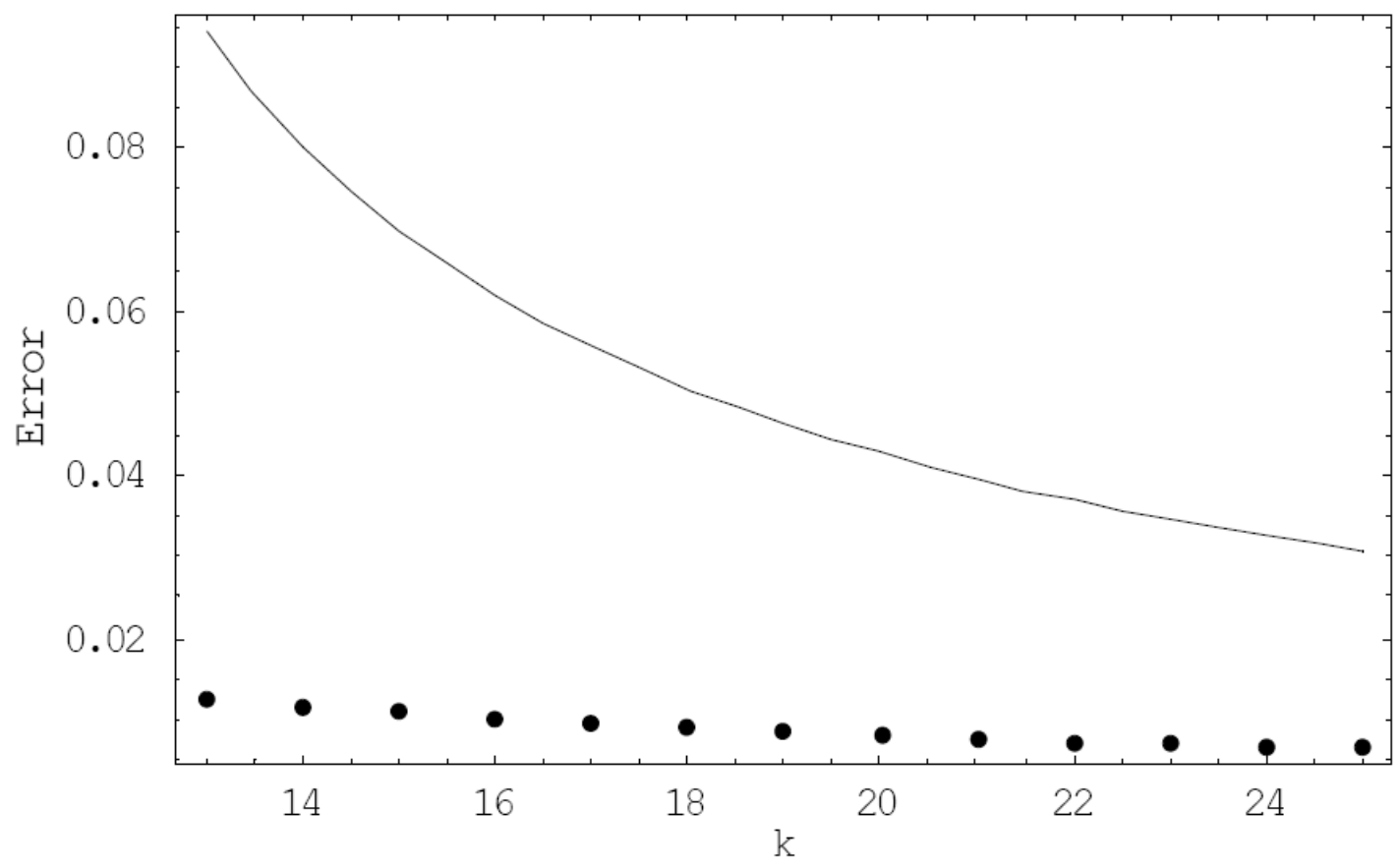

Fig.9 Synchronization error bounds versus the different coupling coefficients $k$. The solid curve and dotted curve represent the estimated synchronization error bound $H$ and the numerical error bound $\|e\|_{\max}$, respectively.

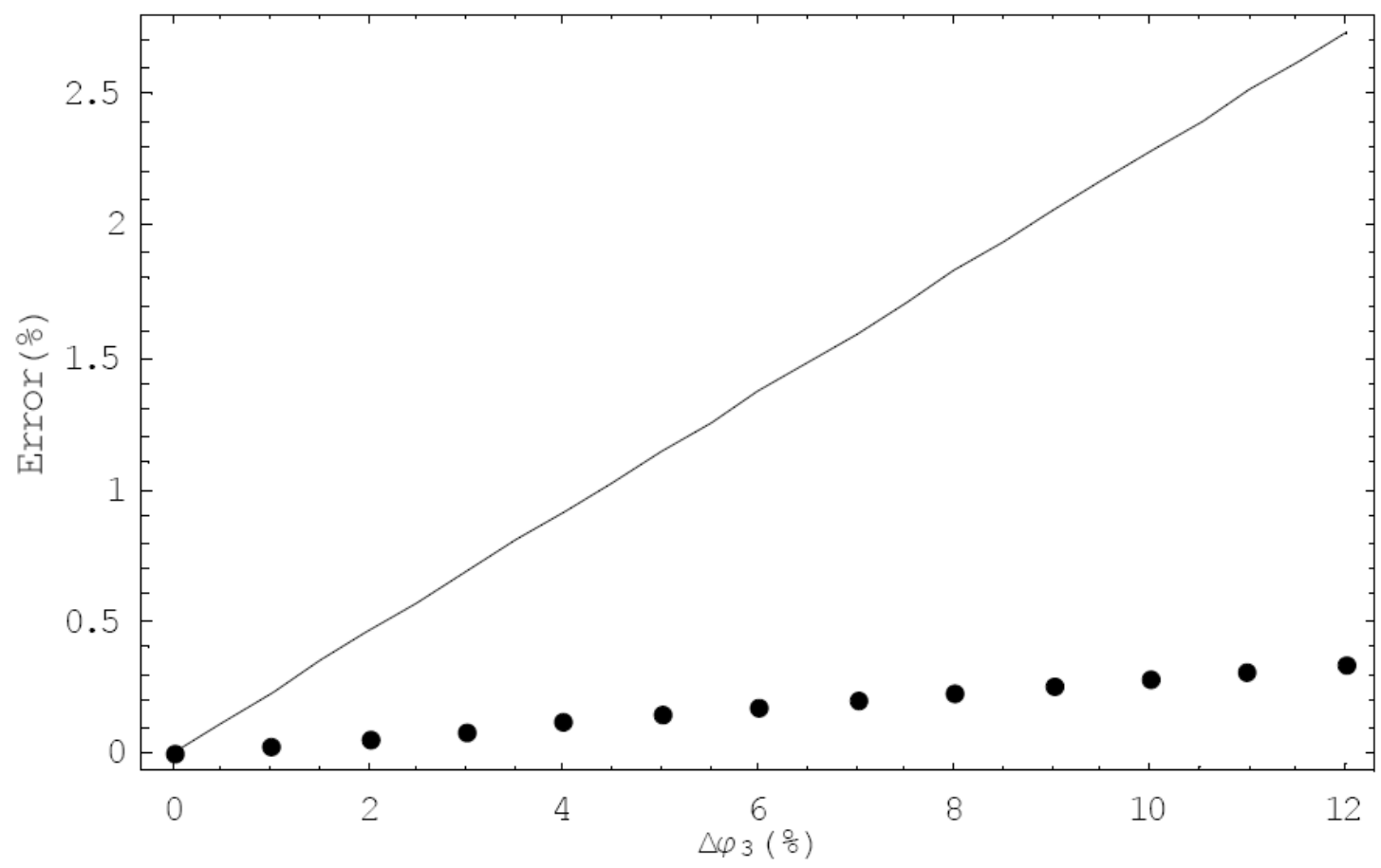

Fig.10 Synchronization error bounds versus the different phase mismatches $\Delta\varphi_3(\%)$. The solid curve and dotted curve represent the estimated synchronization error bound $H(\%)$ and the numerical error bound $\|e\|_{\max}(\%)$, respectively.